\documentclass[aps,prl,10pt,twocolumn,superscriptaddress,longbibliography,nofootinbib]{revtex4-2}

\usepackage[T1]{fontenc}
\usepackage[utf8]{inputenc}
\usepackage{color}
\usepackage{babel}
\usepackage{verbatim}
\usepackage{float}
\usepackage{amsmath,amssymb,amstext,amsfonts}
\usepackage{mathrsfs}
\usepackage{mathtools}
\usepackage{bm}
\usepackage{lipsum}
\usepackage{graphicx}
\usepackage[unicode=true,pdfusetitle,
bookmarks=true,bookmarksnumbered=false,bookmarksopen=false,
breaklinks=true,pdfborder={0 0 1},backref=false,colorlinks=true]
{hyperref}
\usepackage{svg}

\makeatletter

\usepackage{graphicx}
\usepackage{epstopdf}
\usepackage{epsfig}

\newcommand{\ket}[1]{\vert #1 \rangle}

\newcommand{\eV}{{\text{eV}}}
\newcommand{\eVA}{{\text{eV}\cdot \text{\AA}}}
\newcommand{\bk}{\mathbf{k}}
\newcommand{\bq}{\mathbf{q}}

\newcommand{\bd}{\mathbf{d}}

\def\I{\uppercase\expandafter{\romannumeral 1}}
\def\II{\uppercase\expandafter{\romannumeral 2}}
\def\III{{\uppercase\expandafter{\romannumeral 3}}}
\def\IV{{\uppercase\expandafter{\romannumeral 4}}}
\def\V{{\uppercase\expandafter{\romannumeral 5}}}
\def\VI{{\uppercase\expandafter{\romannumeral 6}}}
\def\VII{{\uppercase\expandafter{\romannumeral 7}}}
\def\i{\lowercase\expandafter{\romannumeral 1}}
\def\ii{\lowercase\expandafter{\romannumeral 2}}
\def\iii{{\lowercase\expandafter{\romannumeral 3}}}

\def\a{\mathbf{a}}
\def\b{\mathbf{b}}
\def\p{\mathbf{p}}

\def\k{\mathbf{k}}

\def\bd{\mathbf{d}}

\def\R{\mathbf{R}}

\makeatother

\begin{document}

\title{Gradient-based optimization of non-Abelian fractional quantum states in patterned superlattices}

\author{Yifei Guan}
\affiliation{School of Physical Science and Technology, ShanghaiTech Laboratory for Topological Physics, State Key Laboratory of Quantum Functional Materials, ShanghaiTech University, Shanghai 201210, China}

\author{Lichen Yu}
\affiliation{School of Physical Science and Technology, ShanghaiTech Laboratory for Topological Physics, State Key Laboratory of Quantum Functional Materials, ShanghaiTech University, Shanghai 201210, China}

\author{Zizhuang Liu}
\affiliation{School of Physical Science and Technology, ShanghaiTech Laboratory for Topological Physics, State Key Laboratory of Quantum Functional Materials, ShanghaiTech University, Shanghai 201210, China}

\author{Xin Lu}
\affiliation{School of Physical Science and Technology, ShanghaiTech Laboratory for Topological Physics, State Key Laboratory of Quantum Functional Materials, ShanghaiTech University, Shanghai 201210, China}

\author{Jianpeng Liu}
\email{liujp@shanghaitech.edu.cn}
\affiliation{School of Physical Science and Technology, ShanghaiTech Laboratory for Topological Physics, State Key Laboratory of Quantum Functional Materials, ShanghaiTech University, Shanghai 201210, China}
\affiliation{Liaoning Academy of Materials, Shenyang 110167, China}

\date{\today}

\begin{abstract}
The realization of fractional Chern insulator (FCI) states in moir\'e heterostructures has attracted intense interest in the study of correlated topological states. So far, most experimentally realized FCI states may be interpreted as lattice analogues of Abelian fractional quantum Hall (FQH) states. Realizing non-Abelian FCI states is  an important challenge in the field. Patterned dielectric superlattices provide a versatile platform for engineering  topological flat bands. Such systems offer substantial structural flexibility and tunability, because their lattice patterns, periods, and other structural parameters can all be designed and fabricated. Here, 
we provide a gradient-based optimization workflow to design non-Abelian fractional states in patterned bilayer graphene superlattices. The experimentally relevant structural parameters of the superlattice devices are gradient-optimized to favor a flat Chern band with quantum-geometric properties reminiscent of those of the first excited Landau level. Exact diagonalization calculations at 1/2 filling of the optimized Chern band naturally yield non-Abelian FCIs. We apply this workflow to triangular, honeycomb, and kagome patterned superlattices and find robust non-Abelian FCIs  over a large region of the parameter space spanned by  superlattice constant and vertical potential drop. Our work thus establishes an experimentally feasible framework for exploring non-Abelian FCIs in realistic patterned-superlattice devices. 
\end{abstract}

\maketitle

\textit{Introduction}
By introducing a characteristic length scale much larger than that of atomic lattices, lateral superlattices enable the engineering of topological flat bands in two dimensions \cite{macdonald-pnas11,origin-magic-angle-prl19,song-tbg-prl19,yang-tbg-prx19,jpliu-prb19,zaletel-tbg-2019,zhang-senthil-tbg19,wu-TI-twisted-tmd-prl19,jpliu-prx19,wang-prl22,eslam-tmg-prl22}, which provide a versatile platform for the emergence of novel quantum states \cite{andrei-review-tbg,cao-nature18-supercond,cao-nature18-mott}. Typical examples are the fractional Chern insulators (FCIs)  realized in twisted moir\'e superlattices  \cite{fqah-nature23,fqah-prx23,fqah-optics-xu-nature23,fqah-mak-nature23,xu-mote2-np26,li-mote2-arxiv26,fqah-ju-nature24,lu-hexlayer-arxiv24,ju-eqah-nature25,lu-nature26,ashoori-prx25}, which may be interpreted as zero-magnetic-field, lattice analogues of Abelian fractional quantum Hall (FQH) states \cite{fci-prx11,sheng-fci-nc11,murdy-fci-prl11,wen-kagome-prl11,sarma-flatchern-prl11,cooper-fci-prl09,qi-fqah-prl11}. Some FQH states, such as the celebrated Moore-Read state at 1/2 filling of the first excited Landau level (LL) \cite{moore-read,wilczek-pair-prl91}, may host non-Abelian anyon excitations \cite{nayak-1996}, which have potential applications in topological quantum computation \cite{nayak-rmp08}. Such non-Abelian fractional states have been theoretically studied in lattice models \cite{kapit-prl10,liu-nafci-prb13,wang-fci-prb15,he-prb20} and have recently been proposed in moir\'e flat bands \cite{fu-mote2-prl24,song-mr-prb24,ahn-mr-prb24,xu-mote2-prl25,wang-mote2-prl25,sheng-nafci-nc25,liu-nafci-prl25,niu-sheng-prb25,wu-mote2-prb26}. However, convincing experimental evidence for non-Abelian FCIs has yet to be observed, and obtaining such evidence remains a major challenge in the field.

\begin{figure*}
	\centering
	\includegraphics[width=1\linewidth]{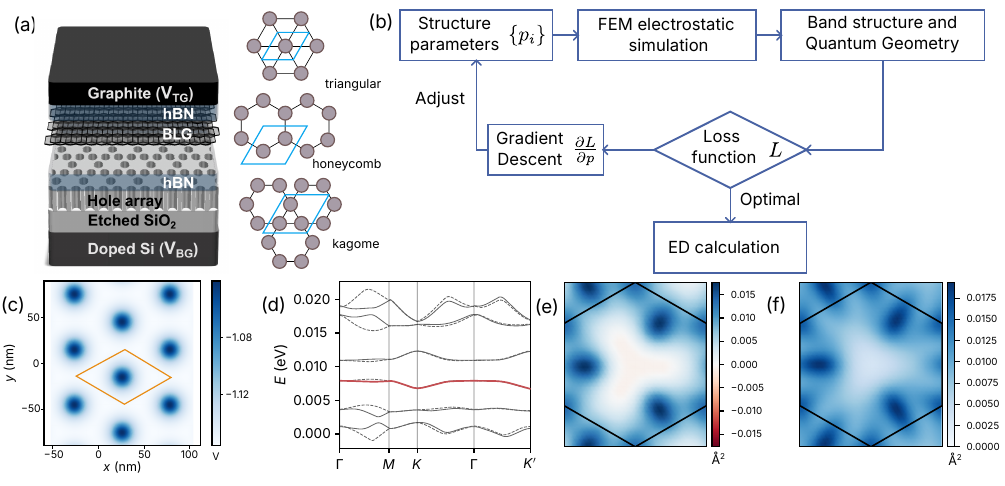}
	\caption{(a) The left panel shows the superlattice device: bilayer graphene is encapsulated by thin hBN layers. A periodically etched SiO$_2$ substrate provides a periodic modulation of the vertical electric potential drop between the top and bottom gates. The right panel shows schematic diagrams of the superlattices with triangular, honeycomb, and kagome patterns. (b) Illustration of the optimization workflow (see text).
	(c) An example of the real-space superlattice potential acting on the bottom graphene layer for a triangular-patterned superlattice with gradient-optimized structural parameters under fixed $\delta V=2\,$V and $L_s=60\,$nm, where the orange line marks one unit cell. (d)--(f) Representative electronic properties of bilayer graphene coupled to the gradient-optimized  potential shown in (c): (d) the band structure, (e) the Berry-curvature distribution, and (f) the distribution of the trace of the quantum metric $\operatorname{tr} g(\mathbf{k})$.}
	\label{fig:fig1}
\end{figure*}

Most of previous  proposals of non-Abelian FCIs are focused on twisted moir\'e systems \cite{fu-mote2-prl24,song-mr-prb24,ahn-mr-prb24,xu-mote2-prl25,wang-mote2-prl25,sheng-nafci-nc25,liu-nafci-prl25,niu-sheng-prb25,wu-mote2-prb26}, while little focus on  topological flat bands emerging from other superlattice platforms. 
Patterned dielectric superlattices, which integrate prepatterned dielectric substrates containing periodic arrays of etched holes with few-layer van der Waals materials \cite{dean-nn18,chen-cp20,dean-nn21,ruiz-nc22,zeng-prl24,du-nanoletter24}, provide a new platform for engineering topological flat bands in two dimensions \cite{cano-bilayer-prl23,lu-nc23,cano-multilayer-prb23,vanderbilt-bilayer-prb24,zhan-patterned-prb25,trithep-semiconductor-prl24,dai-artificial-arxiv24,shi-prl25,cano-prb26}.
In twisted moir\'e systems,  appearance of topological flat bands requires precise control of the twist angle, and it usually suffers from twist-angle disorder \cite{zeldov-disorder-np20} and uncontrolled lattice relaxations \cite{kazmierczak2021strain}. 
Patterned superlattice systems, on the other hand, offer greater structural tunability and variety. In particular, the lattice patterns, periods, and other structural parameters can all be designed and fabricated with high precision using state-of-the-art micro- and nanofabrication techniques \cite{dean-nn18,chen-cp20,dean-nn21,ruiz-nc22,zeng-prl24,du-nanoletter24,shi-prl25}. This flexibility provides an enormous parameter space and structural variety for exploring novel topological quantum states, such as the lattice constant $L_s$, the sublattice pattern, and the depth and radius of each etched hole. It is impractical to search over the entire parameter space. This raises a central question: how can appropriate superlattice devices be designed to realize desired novel quantum states, such as non-Abelian FCI states?

In this work, we introduce a gradient-based workflow that directly optimizes the experimentally relevant structural parameters of the device, such as depth and radius of the etched hole, to realize non-Abelian fractional states in patterned dielectric superlattices coupled to  bilayer graphene (BLG). 
The optimization yields  device structures that give rise to topological flat bands approaching the ideal quantum geometry of the first excited LL, a prerequisite for hosting a non-Abelian Moore-Read state at half filling. This gradient-based optimization framework is applied to triangular, honeycomb, and kagome patterned superlattices coupled to BLG, all of which can induce first-LL-like flat Chern bands. Using many-body energy and particle entanglement spectra obtained by exact diagonalization at half filling of the optimized flat band, we confirm that robust non-Abelian FCI states exist over large regions of parameter space for all three types of superlattices.

\textit{Superlattice device setup and optimization workflow}
The device is built in the layered heterostructure shown in Fig.~\ref{fig:fig1}(a) \cite{supp_info}. 
A voltage drop $\delta V$ applied between the top and bottom gates produces an electrostatic potential in the BLG, which is modulated by the periodically patterned SiO$_2$ substrate. The superlattice potential is determined not only by the gate potential difference $\delta V$ and superlattice constant $L_s$, but also by a set of internal structural parameters $\{p_i\}$ including hole radii, and hole depths etc. 

Figure~\ref{fig:fig1}(b) summarizes the optimization workflow. Starting from a trial structural parameter set $\{p_i\}$ at fixed $\delta V$ and $L_s$, we first simulate the electrostatic potential distribution using the finite-element method (FEM), which produces a superlattice-periodic potential $V_{\rm SL}^{l}(\mathbf r)$ acting on graphene layers $l=1,2$. Within each layer, $V_{\text{SL}}^{l}$ acts as a sublattice-independent scalar potential on the $l$th graphene layer. 
The low-energy electronic structure of bilayer graphene coupled to such a superlattice potential is described by a realistic continuum model
\begin{align}
  H_{\mu}=H_{\mu}^{0}+V_{\rm SL},\;
  \label{eq:ham}
\end{align}
where $H_{\mu}^{0}$ is a realistic low-energy $k\cdot p$ Hamiltonian of BLG \cite{moon-tbg-prb13}, and $\mu=\mp1$ labels the $K/K'$ valleys.
Diagonalizing $H_{\mu}$ in the superlattice Brillouin zone gives the subband structures, Berry curvature $\Omega(\mathbf k)$, and quantum metric $g(\mathbf k)$, which are used to evaluate the loss function for the target band \cite{supp_info}. 

\begin{figure*}
	\centering
	\includegraphics[width=0.8\linewidth]{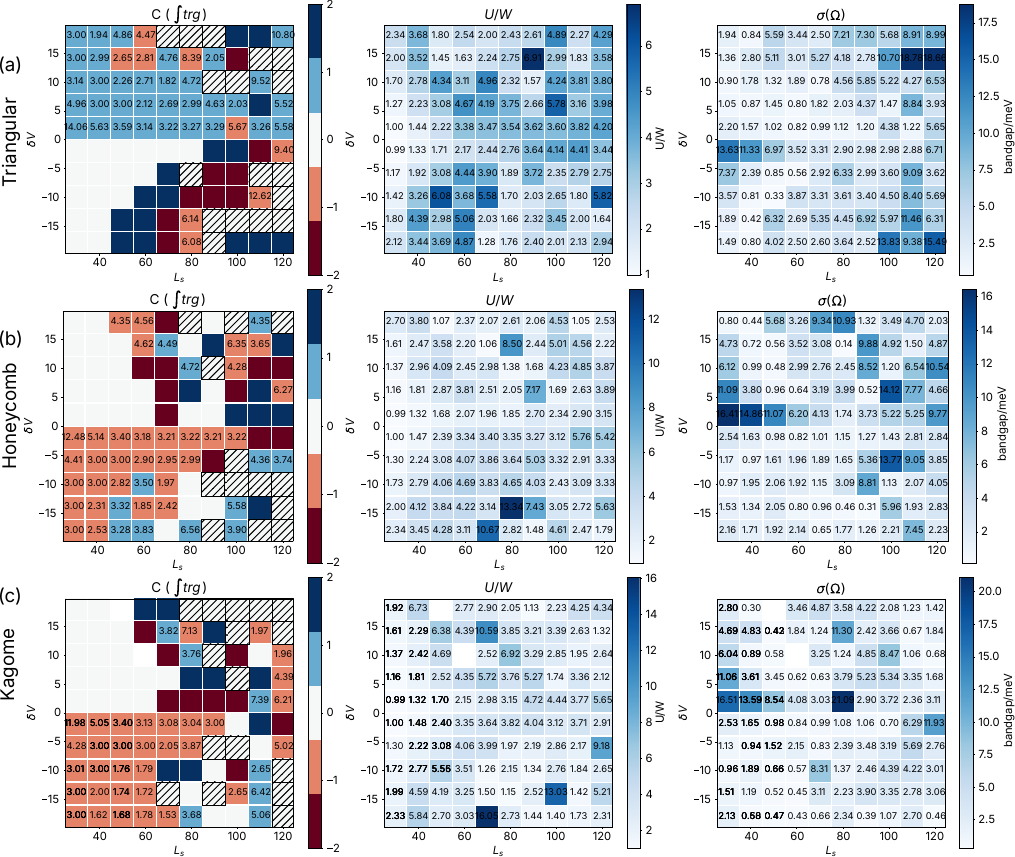}
	\caption{Single-particle phase diagrams of the gradient-optimized superlattice structures for
		(a) triangular, (b) honeycomb, and (c) kagome patterned superlattices in the parameter space spanned by the superlattice constant $L_s$ and gate-voltage difference $\delta V$. The color coding of the subfigures in the first column indicates the Chern numbers of the highest valence bands in the $K$ valley, while the number at each parameter point gives the numerical value of $\int_{\mathbf{k}} \operatorname{tr} g(\mathbf{k})$. The color maps in the second column show the ratio of the characteristic electron-electron Coulomb interaction energy $U=e^2/(4\pi\epsilon_s\epsilon_0 L_s)$ to the bandwidth $W$ of the highest valence band, with the value of $U/W$ marked at each parameter point. The color maps in the third column show the Berry-curvature standard deviations $\sigma(\Omega)$.}
	\label{fig:phasemap}
\end{figure*}

Our optimization is designed to produce an isolated flat Chern band whose quantum geometry resembles that of the first excited LL, because our goal is to revive Moore-Read-type state \cite{moore-read,morf-MR1stLL-prl-1998,stroni-MR1stLL-prl-2010} without magnetic field. The ideal first LL has uniform Berry curvature, Chern number $C=\pm1$, and a uniform quantum metric satisfying $\int_{\mathbf{k}} \operatorname{tr} g(\mathbf{k})=3$ \cite{wang-geometry-prl21,liu-wang-prx25,ledwith-strongcoupling-tbg}, where
$\int_{\mathbf{k}}\equiv \int_{\text{B. Z.}} d^2\mathbf{k}/2\pi$. Therefore, we define the loss function as
\begin{align}
L= \left| \frac{1}{2\pi}\int_{\text{B. Z.}} d^2\mathbf{k} \operatorname{tr}g(\mathbf{k})-3 \right| + \alpha\,\sigma(\Omega),
\label{eq:loss}
\end{align}
where $\sigma(\Omega)$ is the dimensionless standard deviation of the Berry curvature.
The parameter $\alpha$, setting to 0.5, controls the relative weight assigned to the Berry-curvature fluctuations. Although the bandwidth and band gap are not included explicitly in $L$, suppressing Berry-curvature fluctuations also tends to disfavor band touchings, thus favoring an isolated flat band. Note that the loss function is defined for a target band, which is chosen to be the highest valence band (HVB) of the bilayer graphene superlattice.

We evaluate the gradient of the loss function using numerical finite differences:
$\nabla_{\mathbf p}L=\left[\frac{\partial L}{\partial p_1},\frac{\partial L}{\partial p_2},\dots \right]$.
At optimization step $n$, the parameters are updated according to $\mathbf p^{(n+1)}=\mathbf p^{(n)}-\lambda\nabla_{\mathbf p}L$, where $\lambda$ is the learning rate. In this work, we treat  $L_s$ and  $\delta V$ as external parameters, and  optimize the loss function  with respect  to $r_{\text{hole}}$ and $h_{\text{hole}}$ at each $(L_s,\delta V)$ point. 
Representative convergence curves for the loss function $L$, $r_{\text{hole}}$ and $h_{\text{hole}}$ are provided in the Supplemental Material \cite{supp_info}.

\textit{Gradient-optimized single-particle properties} 
Figures~\ref{fig:fig1}(c)--(f) show representative results for a structurally optimized triangular superlattice with fixed $\delta V=2\,$V and $L_s=60\,$nm, for which the hole radius $r_{\text{hole}}$ and depth $h_{\text{hole}}$ are optimized to  $r_{\text{hole}}=13.9\,$nm and $h_{\text{hole}}=20.1\,$nm. Figure~\ref{fig:fig1}(c) presents the optimized real-space superlattice potential acting on the bottom graphene layer.  The resulting HVBs in the $K/K'$ valleys have Chern numbers $\mp1$. As shown in Fig.~\ref{fig:fig1}(d), the target band is  flat and energetically separated from the other bands. Its Berry-curvature distribution in the Brillouin zone is shown in Fig.~\ref{fig:fig1}(e) and has a small dimensionless standard deviation, $\sigma(\Omega)=0.8$. The $\bk$-space distribution of $\operatorname{tr}g(\mathbf{k})$ is shown in Fig.~\ref{fig:fig1}(f), and $\int_{\mathbf{k}}\operatorname{tr}g(\mathbf{k})=3.14$, close to the first-LL value. 

We apply the gradient optimization approach to triangular, honeycomb, and kagome patterned superlattices coupled to bilayer graphene. Figure~\ref{fig:phasemap} presents the single-particle properties of the HVBs of bilayer graphene superlattices with gradient-optimized structures for (a) triangular, (b) honeycomb, and (c) kagome patterns.    The optimized structures exhibit $|C|=1$ flat bands with $\int_{\mathbf{k}} \operatorname{tr} g(\mathbf{k})$ close to 3 over a wide range of $(L_s, \delta V)$ parameter space, as shown in the upper-left region for triangular lattice and lower-left regions for the honeycomb and kagome lattices in the first column of Fig.~\ref{fig:phasemap}. Meanwhile, the Berry-curvature standard deviation $\sigma(\Omega)\lesssim 1$ as shown in the third column. Such single-particle characteristics approach the ideal limit of the first LL. Moreover, as shown in the second column of Fig.~\ref{fig:phasemap}, $U/W$ ranges from 2 to 5 over an extended parameter window, placing the system in the correlation-dominated regime in which FCI phases can be stabilized.

\begin{figure}
	\centering
	\includegraphics[width=1\linewidth]{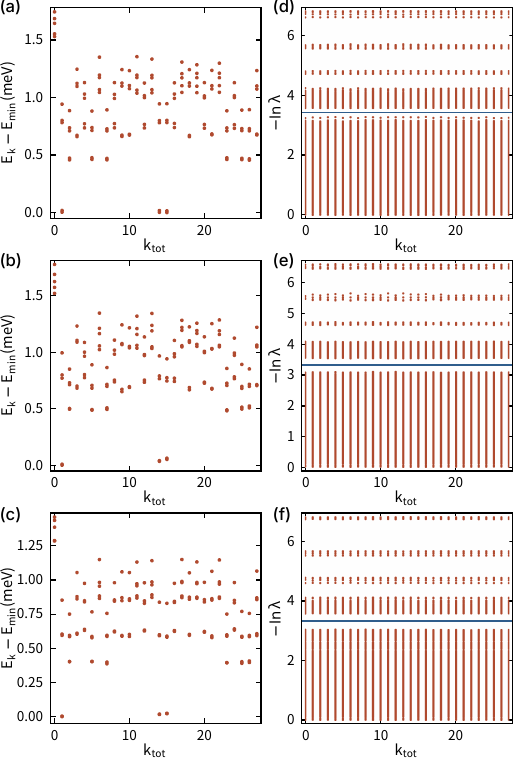}
	\caption{The many-body spectrum and PES of the Moore-Read FCI  phase, calculated with $N_k = 28, N_e = 14$. (a)-(c) show the energy spectra for (a) triangular superlattice with $L_s=40\,\text{nm}$, $\delta V=18\,\text{V}$; (b) honeycomb superlattice with $L_s=40\,\text{nm}$, $\delta V=-14\,\text{V}$; and (c) kagome  superlattice with  $L_s=50\,\text{nm}$, $\delta V=-14\,\text{V}$. (d)-(f) present the corresponding  PES  calculated by a partition $N_A = 4, N_B =10$ of the 14 electrons.}
	\label{fig:fig3-edpes}
\end{figure}

\textit{Non-Abelian FCI states}
To explore the many-body electronic states, we include the  electron-electron Coulomb interactions projected onto the HVB subspace of a single valley-spin flavor, denoted as $V_{\text{int}}^{\text{proj}}$ \cite{supp_info}. The single-particle HVB energy combined with band-projected interaction $V_{\text{int}}^{\text{proj}}$ gives the many-body Hamiltonian for exact diagonalizations.

\begin{figure*}
	\centering
	\includegraphics[width=1\linewidth]{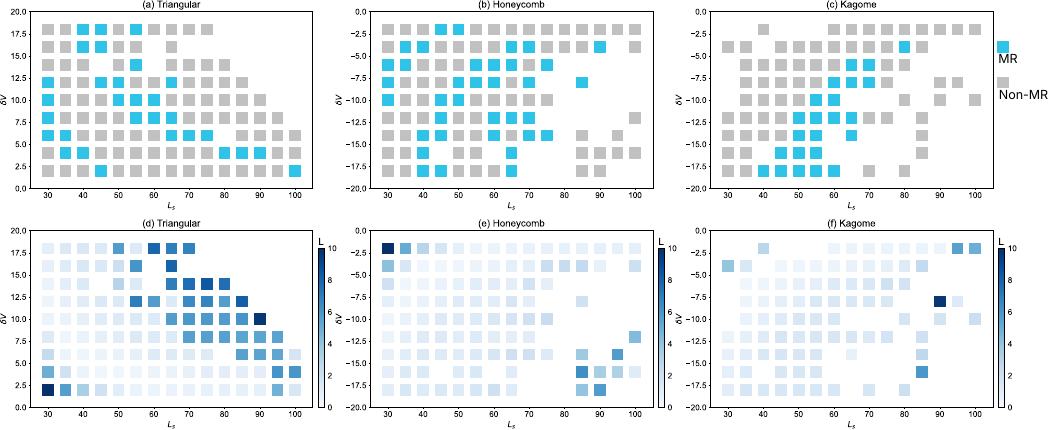}
	\caption{Many-body phase diagrams of the HVB at 1/2 filling in the $(L_s,\delta V)$ parameter space for (a) triangular, (b) honeycomb, and (c) kagome patterned superlattices coupled to bilayer graphene. Moore-Read FCI states are colored light blue, the gray blocks denote many-body states that are not Moore-Read FCIs, and the white regions indicate parameter points at which the Chern number of the HVB is not $\pm1$ and the many-body ground state was therefore not calculated. (d)--(f) show the converged values of the loss function defined in Eq.~\eqref{eq:loss} for the triangular, honeycomb, and kagome superlattices, respectively.}
	\label{fig:fig4-edmap}
\end{figure*}

Many-body spectra are obtained by exact diagonalization (ED) at 1/2 fillings of two torus clusters with system sizes $N_s=28$ and $N_s=26$, respectively, using geometries specified by tilted reciprocal-lattice vectors (see Supplemental Material). 
Figures~\ref{fig:fig3-edpes}(a)--(c) show representative ED energy spectra for triangular, honeycomb, and kagome patterned superlattices with $N_s=28$ at half filling ($N_e/N_s=1/2$). A key finite-size diagnostic of a Moore-Read-type FCI state is the presence of a quasi-degenerate ground-state manifold, and the total momenta of these ground states should be consistent with a generalized Pauli principle \cite{bernevig-prb12,fci-prx11}. For a Moore-Read-type FCI state on a torus, the $(2,4)$-admissible generalized Pauli principle leads to six quasi-degenerate ground states for  even $N_e$ and two for odd $N_e$.
Consistent with this counting, the $N_s=28$, $N_e=14$ spectra exhibit a sixfold quasi-degenerate ground-state manifold distributed among three total-momentum sectors, with two states in each sector, as shown in Figs.~\ref{fig:fig3-edpes}(a)--(c). The $N_s=26$, $N_e=13$ ED calculations yield the expected twofold ground-state manifold \cite{supp_info}. Both the ground-state counting and momentum-sector distributions agree precisely with those expected for  Moore-Read state on a torus. Notably, for the representative optimized superlattice structures presented in Figs.~\ref{fig:fig3-edpes}(a)--(c), the many-body gaps $\Delta\sim 0.362 \text{--} 0.437\,\mathrm{meV}$, while the energy spread of the ground-state manifold is $\delta E_g\sim 0.018 \text{--}0.063\,\mathrm{meV}$.

To further confirm the topological nature, we calculate the particle entanglement spectrum (PES) of the quasi-degenerate ground-state manifold \cite{li-haldane-prb08,bernevig-pes-prb11,Sterdyniak2012a}. A topologically ordered ground state is typically characterized by a well-defined entanglement gap, which separates the low-lying PES manifold from higher entanglement levels. The number and momentum-sector distributions of the low-lying entanglement levels below the gap can then be compared with the quasihole counting predicted for a candidate topological order. Figures~\ref{fig:fig3-edpes}(d)--(f) show representative PES for the triangular, honeycomb, and kagome superlattices. For the $N_s=28$, $N_e=14$ clusters, we use a particle cut with $N_A=4$ and $N_B=N_e-N_A=10$. In all three cases, a clear entanglement gap separates 18571 low-lying levels from the remaining higher levels distributed at correct momentum sectors, which agree exactly with the quasihole counting expected for  Moore-Read state. We have also calculated the PES for the $N_s=26$, $N_e=13$ clusters with a particle cut of $N_A=3$ and $N_B=10$. These spectra also exhibit a clear entanglement gap, with 2522 low-lying entanglement levels below the gap, consistent with the Moore-Read quasihole counting \cite{supp_info}. Together with the quasi-degenerate ground-state manifold and its characteristic momentum-sector distribution, this PES counting provides strong evidence for Moore-Read-type non-Abelian topological order, ruling out possible charge-density-wave interpretation \cite{liu-nafci-prl25}.  It is worth noting that the ground states at the three $(L_s,\delta V)$ parameter points shown in Fig.~\ref{fig:fig3-edpes} become non-Abelian FCIs only after optimization of the loss. Without optimization, the ground-state degeneracy and PES counting are inconsistent with FCI topological order, as shown in the Supplemental Material.


\textit{Many-body phase diagrams}
Figures~\ref{fig:fig4-edmap}(a)--(c) present the many-body phase diagrams of the optimized triangular, honeycomb, and kagome superlattices at half filling of the HVB. Moore-Read-type FCIs occupy an extended and relatively connected region for the triangular superlattice, whereas they are more interspersed with non-Moore-Read states in the honeycomb and kagome cases. It is noteworthy that the Moore-Read states occur for positive $\delta V$ in the triangular superlattice but for negative $\delta V$ in the honeycomb and kagome superlattices. This sign difference can be understood from the difference in leading Fourier components of the superlattice potential due to sublattice structure factors \cite{supp_info}. 

Comparison with the converged loss functions in Figs.~\ref{fig:fig4-edmap}(d)--(f) reveals a moderate correlation between the single-particle loss and the occurrence of Moore-Read order for all three superlattices. 
Specifically, the average loss values of the Moore-Read FCI states are $1.42$, $1.20$, and $1.64$ for the triangular, honeycomb, and kagome patterns, respectively. In comparison, the corresponding average losses of the non-Moore-Read states are $3.55$, $2.43$, and $2.72$. 
However, some non-Moore-Read states also occur at parameter points with low loss.
These results demonstrate that minimizing the loss function indeed effectively favors the first-LL-like quantum geometry conducive to non-Abelian topological order. The  exceptions further indicate that the loss is a heuristic descriptor rather than a sufficient criterion, because other ingredients, such as band dispersion, also influence the many-body ground states.
In Supplemental Materials we also present the many-body phase diagrams at 1/2 filling of the gradient optimized HVB of BLG continuum model with only nearest neighbor interlayer hopping, dubbed as ``chiral model". It turns out that Moore-Read FCI states emerge in a larger region for such chiral BLG model, and have more evident correlations with the amplitude of loss function.

\textit{Discussion}
In summary, we have developed a gradient-based approach for designing non-Abelian fractional topological states in patterned dielectric superlattices. Applying this approach to bilayer-graphene-based systems, we optimize the structural parameters of triangular, honeycomb, and kagome superlattices to obtain isolated flat Chern bands whose quantum geometry approaches that of the first excited LL. At half filling, exact diagonalization calculations reveal the characteristic ground-state degeneracy and PES counting of the Moore-Read fractional states. Such non-Abelian FCIs occur over extended regions of parameter space for all three superlattice patterns, with their stability strongly correlated with the optimized single-particle loss function defined by quantum geometry. These results demonstrate that structurally optimized patterned dielectric superlattices provide a realistic and tunable platform for realizing non-Abelian topological order without an external magnetic field.
 
The device-level design strategy introduced in this work differs from previous gradient-based searches for FCIs in idealized lattice models, which explore Hamiltonian space by directly varying hopping and interaction amplitudes \cite{gradient-fci-arxiv26}. 
Our framework  translates the target quantum geometry into concrete device-design guidelines for realizing non-Abelian topological states in patterned-superlattice experiments. Although demonstrated here for bilayer graphene, the framework  can be readily extended to patterned dielectric superlattices coupled to a broad range of two-dimensional van der Waals materials and multilayer systems, including transition-metal dichalcogenides and rhombohedral multilayer graphene. Our work therefore provides a systematic route from experimentally accessible device parameters to the realization of novel correlated topological matter.

\textit{Acknowledgments}
This work is supported by the National Key Research and Development Program of China (grant nos. 2024YFA1410400 and 2022YFA1604400/03), the National Natural Science Foundation of China (grant nos. 12504193, 12550403 and 12404221), the Strategic Priority Research Program of the Chinese Academy of Sciences (grant no. XDB1710000), the Shanghai Science and Technology Innovation Action Plan (grant no. 24LZ1401100) and the Quantum Science and Technology-National Science and Technology Major Project (grant no. 2025ZD0300500).

\bibliography{reference}

\clearpage
\onecolumngrid

\setcounter{section}{0}
\setcounter{subsection}{0}
\setcounter{subsubsection}{0}
\setcounter{equation}{0}
\setcounter{figure}{0}
\setcounter{table}{0}
\renewcommand{\thesection}{S\arabic{section}}
\renewcommand{\thesubsection}{\thesection.\arabic{subsection}}
\renewcommand{\thesubsubsection}{\thesubsection.\arabic{subsubsection}}
\renewcommand{\theequation}{S\arabic{equation}}
\renewcommand{\thefigure}{S\arabic{figure}}
\renewcommand{\thetable}{S\arabic{table}}
\renewcommand{\theHsection}{supp.section.\arabic{section}}
\renewcommand{\theHsubsection}{supp.subsection.\arabic{section}.\arabic{subsection}}
\renewcommand{\theHsubsubsection}{supp.subsubsection.\arabic{section}.\arabic{subsection}.\arabic{subsubsection}}
\renewcommand{\theHequation}{supp.equation.\arabic{equation}}
\renewcommand{\theHfigure}{supp.figure.\arabic{figure}}
\renewcommand{\theHtable}{supp.table.\arabic{table}}

\begin{center}
{\large\bfseries Supplemental Material for ``Gradient-based optimization of non-Abelian fractional quantum states in patterned superlattices''}
\end{center}
\vspace{1em}

\section{Non-interacting continuum model}

We derive a continuum model of bilayer graphene subject to a patterned superlattice potential. The full single-particle Hamiltonian is expressed as $H^{\mu}=H^{0,\mu}+V_{\mathrm{SL}}$, where $V_{\mathrm{SL}}$ is the superlattice potential acting on bilayer graphene and the noninteracting part $H^{0,\mu}$ is the $\k\cdot\p$ Hamiltonian for valley $\mu$, derived from the Slater-Koster tight-binding model in Refs.~\onlinecite{Moon-Koshino-prb2014,moon-tbg-prb13}. The tight-binding Hamiltonian for $p_z$-like Wannier orbitals reads
\begin{equation}
  H^{0,\mu}=\sum_{i l\alpha,j l^{\prime}\alpha^{\prime}}
  -t(\R_i+\bm{\tau}_{\alpha}+l d_0\mathbf{e}_z
      -\R_j-\bm{\tau}_{\alpha^{\prime}}-l^{\prime} d_0\mathbf{e}_z)
  \,\hat{c}^{\dagger}_{i l\alpha}\hat{c}_{j l^{\prime}\alpha^{\prime}},
\end{equation}
where $i,j$ label unit cells, $l,l^{\prime}$ label layers, and $\alpha,\alpha^{\prime}$ label sublattices. Here, $d_0=0.335\,$nm is the interlayer distance, and $\mathbf{e}_z$ is the out-of-plane unit vector.
The primitive lattice vectors of graphene are $\a_1=a(1,0)$ and $\a_2=a(1/2,\sqrt{3}/2)$, with $a=2.46\,\text{\AA}$, and the reciprocal vectors are $\b_1=2\pi/a\,(1,-1/\sqrt{3})$ and $\b_2=2\pi/a\,(0,2/\sqrt{3})$, satisfying $\a_i\cdot\b_j=2\pi\delta_{ij}$. The two sublattice sites are located at $\bm{\tau}_a=a(0,-1/\sqrt{3})$ and $\bm{\tau}_b=(0,0)$, and consecutive layers are shifted by $a(0,-1/\sqrt{3})$ in Bernal-stacked bilayer graphene. The Dirac points are located at $\mathbf{K}^{\mu}=-\mu\,4\pi/(3a)\,(1,0)$, with valley index $\mu=\pm1$. The hopping amplitude follows the Slater-Koster form
\begin{align}
  -t(\bd) &= V_{pp\pi}\bigl[1-(\bd\cdot\mathbf{e}_z/d)^2\bigr]
            +V_{pp\sigma}\,(\bd\cdot\mathbf{e}_z/d)^2,\\
  V_{pp\pi} &= V_{pp\pi}^0\;\exp\!\Bigl(-\frac{|\bd|-a/\sqrt{3}}{r_0}\Bigr),\nonumber\\
  V_{pp\sigma} &= V_{pp\sigma}^0\;\exp\!\Bigl(-\frac{|\bd|-d_0}{r_0}\Bigr),\nonumber
\end{align}
where $d=|\bd|$, $V_{pp\pi}^0=-2.7\,\eV$, $V_{pp\sigma}^0=0.48\,\eV$, and $r_0=0.184a$~\cite{moon-tbg-prb13,Moon-Koshino-prb2014}. This realistic Slater-Koster tight-binding model agrees well with band structures obtained from density functional theory and is widely used in the literature.

Fourier transforming the Hamiltonian and expanding it around $\mathbf{K}^{\mu}$ yield the $\k\cdot\p$ model, whose intralayer and interlayer blocks are
\begin{align}
  h_{\mathrm{intra}}^{0,\mu} &= -\hbar v_F^0\,\k\cdot\bm{\sigma}^{\mu},  \\
  h_{\mathrm{inter}}^{0,\mu} &=
  \begin{pmatrix}
    \hbar v_{\perp}(\mu k_x+ik_y) & t_{\perp} \\[2pt]
    \hbar v_{\perp}(\mu k_x-ik_y) & \hbar v_{\perp}(\mu k_x+ik_y)
  \end{pmatrix},
  \label{eq:hintra-inter}
\end{align}
where $\bm{\sigma}^{\mu}=(\mu\sigma_x,\sigma_y)$ are Pauli matrices in sublattice space and $\k$ is measured from $\mathbf{K}^{\mu}$. The Slater-Koster parameters are $\hbar v_F^0=5.253\,\eVA$, $\hbar v_{\perp}=0.335\,\eVA$, and $t_{\perp}=0.34\,\eV$. The bilayer Hamiltonian $H^{0,\mu}$ consists of $h_{\mathrm{intra}}^{0,\mu}$ on each layer and $h_{\mathrm{inter}}^{0,\mu}$ coupling the two layers.

\section{Optimization workflow}

\paragraph{Single-particle Hamiltonian.} The full single-particle Hamiltonian
of the system reads
\begin{equation}
  H^{\mu} = H^{0,\mu} + V_{\mathrm{SL}}(\{p_i\}),
  \label{eq:fullH}
\end{equation}
where $H^{0,\mu}$ is the $\mathbf{k}\cdot\mathbf{p}$ Hamiltonian of bilayer
graphene derived above, and $V_{\mathrm{SL}}(\{p_i\})$ is the
periodic superlattice potential fully determined by the geometric parameters
$\{p_i\}$. Note that $V_{\mathrm{SL}}(\{p_i\})$ is layer dependent, but within each layer $l$, it acts as a sublattice-independent scalar potential $V_{\mathrm{SL}}^{l}(\mathbf{r};\{p_i\})$. The electronic band structures and quantum-geometric properties  follow from diagonalizing
Eq.~\eqref{eq:fullH}.

For the patterned superlattices studied in this work, $\{p_i\}$
reduces to the hole radius $r_{\mathrm{hole}}$ and depth $h_{\mathrm{hole}}$,
i.e., $\mathbf{p} = [r_{\mathrm{hole}}, h_{\mathrm{hole}}]$, while the triangular,
honeycomb, and kagome superlattices correspond to different spatial
arrangements of the holes in the substrate. The superlattice constant $L_s$ and gate-voltage difference $\delta V$ are treated as fixed parameters, and at each $(L_s,\delta V)$ parameter point, we optimize the loss function defined in Eq.~(2) of the main text with respect to $r_{\text{hole}}$ and $h_{\text{hole}}$. The explicit form of
$V_{\mathrm{SL}}$ is obtained from electrostatic simulations of the device using finite-element method. 

The superlattice potential can be expanded into a Fourier series, $V_{\rm SL}^{l}(\mathbf r)= \sum_{\mathbf Q} V_{\mathbf Q}^{l} e^{i\mathbf Q\cdot\mathbf r}$, where $\mathbf Q$ denotes a reciprocal lattice vector of the superlattice. We retain Fourier components through the seventh order, which reproduce the FEM potential with an error below $0.01\%$. For superlattices consisted of multiple identical etched holes in one unit cell forming  sublattice patterns located at $\{\mathbf{\tau}_{\alpha}\}$ ($\alpha$ is sublattice index), $V_{\mathbf{Q}}^{l}$ can be expressed as: $V_{\mathbf{Q}}^{l}=V_0(\mathbf{Q}) \sum_{\alpha}  e^{-i\mathbf{\tau}_{\alpha}\cdot\mathbf{Q}}$, with $V_0(\mathbf{Q})\sim \delta V$. 
For the first-harmonic potential, the sublattice structure factor satisfies $S(\mathbf{Q})=\sum_{\alpha}  e^{-i\mathbf{\tau}_{\alpha}\cdot\mathbf{Q}}=-1$ for honeycomb and kagome patterns, while the triangular pattern has trivial structure factor $S(\mathbf{Q})$.  This minus sign explains why the non-Abelian topological states emerge for positive $\delta V$ for triangular superlattice, while for negative $\delta V$ for honeycomb and kagome superlattices.

\subsection{Device setup}

The device is placed between a graphite top gate and a doped-silicon bottom gate.
\begin{figure}
  \centering
  \includegraphics[width=0.5\linewidth]{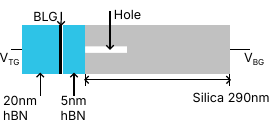}
  \caption{Dimensions of the superlattice device. The electrostatic model consists of an hBN/BLG/hBN stack placed on an etched SiO$_2$ (silica) substrate. The hBN layer adjacent to the top gate is 20\,nm thick, the hBN layer below the bilayer graphene is 5\,nm thick, and the SiO$_2$ substrate is 290\,nm thick.}
  \label{fig:device}
\end{figure}
Figure~\ref{fig:device} shows a side view and the dimensions of the patterned superlattice device.
The SiO$_2$ substrate is 290\,nm thick, and the etched holes have typical depths of 10--30\,nm.
A 5\,nm-thick hBN spacer is placed between the SiO$_2$ substrate and bilayer graphene.
A 20\,nm-thick hBN dielectric layer separates bilayer graphene from the graphite top gate.
The superlattice is formed by the etched holes in the SiO$_2$ substrate.
The superlattice geometry is characterized by the superlattice constant $L_s$, hole depth $h_{\text{hole}}$, hole radius $r_{\text{hole}}$, and spatial arrangement of the holes. For each fixed lattice pattern, $L_s$, and $\delta V$, only $r_{\text{hole}}$ and $h_{\text{hole}}$ are optimized. In all optimizations, the initial hole depth is set to $h_{\text{hole}}=20\,$nm, and the initial hole radius is set to $r_{\text{hole}}=0.15L_s$.

\subsection{Quantum geometry and loss function}

The relevant single-particle
descriptors are constructed from the quantum geometric tensor defined for the periodic part of the Bloch state $\vert u_{\bk}\rangle$,
\begin{align}
  \mathcal{G}_{ij}(\bk) =
  \langle \partial_{k_i} u_{\bk} \mid
  (1 - \vert u_{\bk}\rangle\langle u_{\bk}\vert) \mid
  \partial_{k_j} u_{\bk} \rangle,
\end{align}
whose real part defines the quantum metric, $g_{ij}=\operatorname{Re}
\mathcal{G}_{ij}$, and whose imaginary part gives the Berry curvature,
$\mathcal{F}_{ij}=-2\operatorname{Im}\mathcal{G}_{ij}$. We denote $\Omega(\bk)=\mathcal{F}_{xy}(\bk)$. For the first excited
Landau level (first LL), the trace of the metric satisfies $\int
\operatorname{tr}g\,d^2\bk/(2\pi)=3$, and the Berry curvature is uniform with
Chern number $\vert C\vert=1$.

To target Moore-Read physics, we therefore require the optimized flat band to
resemble the first LL. This motivates the loss function
\begin{align}
  L =
  \left| \frac{1}{2\pi}\int_{\text{B. Z.}} d^2\bk \operatorname{tr} g - 3 \right|
  + \alpha \, \sigma(\Omega),
\end{align}
where $\sigma(\Omega)$ is the dimensionless standard deviation of the Berry curvature and
the parameter $\alpha$ controls the relative weight assigned to the Berry-curvature fluctuations. We set
$\alpha=0.5$. We note that an explicit bandwidth constraint is not needed: the
Berry-curvature-fluctuation penalty tends to suppress band touchings.

\subsection{Gradient descent optimization}

We optimize $L$ via gradient descent
with finite differences:
\begin{align}
  \frac{\partial L}{\partial p_i}
  \approx
  \frac{L(p_1, \dots, p_i + \delta, \dots)
    - L(p_1, \dots, p_i, \dots)}{\delta},
\end{align}
where the gradient vector is assembled as
$\nabla L=[\partial L/\partial p_1,\,
\partial L/\partial p_2,\dots]^{T}$. Each evaluation of
$\partial L/\partial p_i$ requires a finite-element electrostatic
simulation followed by a band-structure calculation. In the optimization,
each parameter is linearly mapped to the domain $[0,1]$ to normalize the step size across parameters of different physical dimensions.
For the triangular, honeycomb, and kagome lattices, the initial parameter values are set to $r_{\text{hole}}=0.15L_s$ and $h_{\text{hole}}=20\,\text{nm}$.
\begin{figure}
  \centering
  \includegraphics[width=1\linewidth]{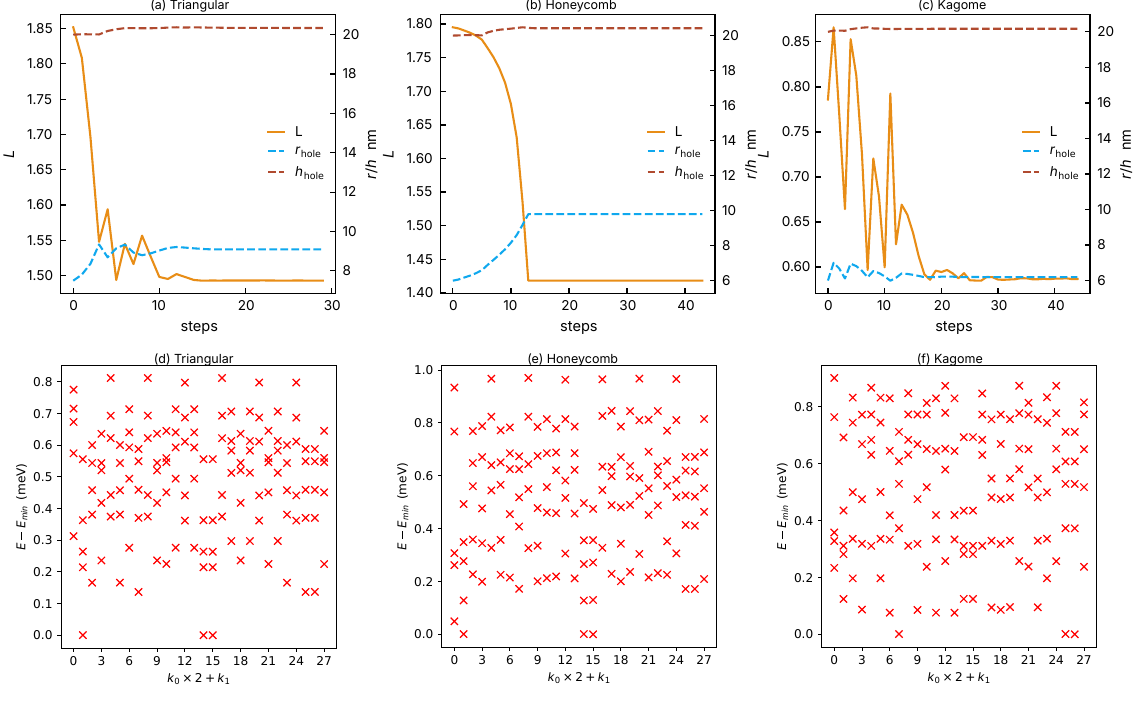}
  \caption{Gradient-descent convergence for representative (a) triangular, (b) honeycomb, and (c) kagome superlattices. The $(L_s,\delta V)$ parameters correspond to those used in Fig.~3 of the main text. Panels (d)--(f) show the corresponding many-body energy spectra at 1/2 filling of the highest valence bands for the unoptimized initial structures with $r_{\text{hole}}=0.15L_s$ and $h_{\text{hole}}=20\,$nm. The ground-state degeneracies are inconsistent with those expected for a Moore-Read state. Moore-Read FCI states emerge only after structural optimization, as shown in Fig.~3 of the main text.}
  \label{fig:gds}
\end{figure}
Figures~\ref{fig:gds}(a)--(c) show representative convergence curves of $r_{\text{hole}}$, $h_{\text{hole}}$, and the loss $L$ as functions of the gradient-descent iteration for the three cases presented in Fig.~3 of the main text.
We also calculate the many-body energy spectra by exact diagonalization of the band-projected many-body Hamiltonian (see the following section) using the initial structural parameters $r_{\text{hole}}=0.15L_s$ and $h_{\text{hole}}=20\,$nm. For the three cases shown in Figs.~\ref{fig:gds}(a)--(c), the ground-state degeneracies are clearly inconsistent with those expected for a Moore-Read FCI, as shown in Figs.~\ref{fig:gds}(d)--(f). Only after structural optimization do the many-body ground states become Moore-Read states, as shown in Fig.~3 of the main text.

\section{Exact diagonalization and particle entanglement spectra}

\subsection{Details of band-projected interaction Hamiltonian}

We first express the electron-electron Coulomb interaction in the band basis. The dominant intravalley interaction reads \cite{lee-tdbg-nc19,zhang-tbmg-prl22}
\begin{align}
  \hat{V}^{\rm{proj}}_{\text{int}}&=\frac{1}{2N_s}\sum_{\bk \bk'\bq}\sum_{\substack{\mu\mu' \\ \sigma\sigma'\\l l'}}\sum_{\substack{nm\\ n'm'}} \left(\sum_{\mathbf{Q}}\,V(\mathbf{Q}+\bq)\,\Lambda^{\mu l,\mu' l'}_{nm,n'm'}(\bk,\bk',\bq,\mathbf{Q})\right) \nonumber \\
  &\times \hat{c}^{\dagger}_{\sigma\mu,n}(\bk+\bq) \hat{c}^{\dagger}_{\sigma'\mu',n'}(\bk'-\bq) \hat{c}_{\sigma'\mu',m'}(\bk') \hat{c}_{\sigma\mu,m}(\bk),
  \label{eq:Hintra-band}
\end{align}
where $N_s$ is the total number of superlattice unit cells in the system, $\hat{c}_{\sigma\mu,n}(\bk)$ annihilates an electron with spin $\sigma$ in the $n$th Bloch band of valley $\mu$ at $\bk$, and the form factor $\Lambda^{\mu l,\mu' l'}_{nm,n'm'}$ reads
\begin{equation}
  \Lambda^{\mu l,\mu' l'}_{nm,n'm'}(\bk,\bk',\bq,\mathbf{Q})
  =\sum_{\alpha\alpha'\mathbf{G}\mathbf{G}'}C^*_{\mu l \alpha\mathbf{G}+\mathbf{Q},n}(\bk+\bq) C^*_{\mu'l'\alpha'\mathbf{G}'-\mathbf{Q},n'}(\bk'-\bq)C_{\mu'l'\alpha'\mathbf{G}',m'}(\bk')C_{\mu l \alpha\mathbf{G},m}(\bk),
\end{equation}
where $C_{\mu l \alpha \mathbf{G},n}(\bk)$ is the wave-function coefficient of the $n$th Bloch eigenstate at $\bk$ in valley $\mu$ in the original $\ket{\sigma,\mu,l,\alpha,\mathbf{G};\bk}$ basis. Here, $\mu$ and $\sigma$ label the valley and spin, $l$ labels the layer, $\alpha$ labels the sublattice, and $\mathbf{G}$ and $\mathbf{Q}$ denote reciprocal lattice vectors of the superlattice. The quantities $n,m,n'$, and $m'$ are band indices, while $\bk,\bk'$, and $\bq$ are wave vectors in the superlattice Brillouin zone.

To model the screening effects of the Coulomb interactions in the superlattice, we use a layer-independent screened Coulomb potential in momentum space,
\begin{equation}
  V(\bq)=\frac{e^2}{2\Omega_0\epsilon_s\epsilon_0\sqrt{|\bq|^2+\kappa^2}},
  \label{eq:Vq}
\end{equation}
where $\Omega_0$ is the area of a superlattice primitive cell, $\kappa=(120\,\mathrm{nm})^{-1}$ is the inverse screening length, and $\epsilon_s=4$ is the relative dielectric constant of hBN.

In our numerical calculations, we project the layer-independent interaction $V(\bq)$ onto a single spin-valley flavor and a single target band, for which Eq.~\eqref{eq:Hintra-band} reduces to
\begin{align}
  \hat{V}^{\rm{int}}_{\text{proj}}&=\frac{1}{2N_s}\sum_{\bk \bk'\bq} \left(\sum_{\mathbf{Q}}\,V(\mathbf{Q}+\bq)\,\Lambda^{\mu}(\bk,\bk',\bq,\mathbf{Q})\right) \nonumber \\
  &\times \hat{c}^{\dagger}(\bk+\bq) \hat{c}^{\dagger}(\bk'-\bq) \hat{c}(\bk') \hat{c}(\bk),
  \label{eq:Hintra-band-single}
\end{align}
with the form factor
\begin{equation}
  \Lambda^{\mu}(\bk,\bk',\bq,\mathbf{Q})
  =\sum_{l l'}\sum_{\alpha\alpha'\mathbf{G}\mathbf{G}'}C^*_{\mu l \alpha\mathbf{G}+\mathbf{Q}}(\bk+\bq) C^*_{\mu l'\alpha'\mathbf{G}'-\mathbf{Q}}(\bk'-\bq)C_{\mu l'\alpha'\mathbf{G}'}(\bk')C_{\mu l \alpha\mathbf{G}}(\bk),
\end{equation}
where $\hat{c}(\bk)$ annihilates an electron in the target band at $\bk$, and $C_{\mu l\alpha\mathbf{G}}(\bk)$ is the corresponding wave-function coefficient in the $\ket{\mu,l,\alpha,\mathbf{G};\bk}$ basis. The projected interaction is independent of spin index due to the negligible spin-orbit coupling in graphene. Combining the single-particle band energy with the band-projected interaction in Eq.~\eqref{eq:Hintra-band-single} gives the many-body Hamiltonian used in the exact diagonalization (ED) calculations. 

\subsection{Tilted reciprocal lattices}
The ED calculations are performed on a tilted momentum mesh
in reciprocal space to better sample the Brillouin zone \cite{repellin-prb14}.
This construction allows us to use a wider variety of system sizes and mesh geometries, including the $N_s=26$ and $N_s=28$ momentum meshes used below, and hence to examine the even-odd particle-number dependence of the ground-state degeneracy, which is
essential for identifying the Moore-Read FCI state.

\begin{figure}
  \centering
  \includegraphics[width=0.5\linewidth]{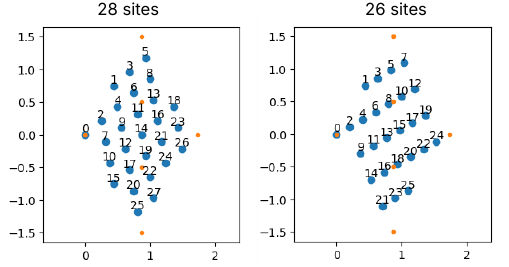}
  \caption{Tilted momentum meshes with 26 and 28 points in reciprocal space.}
  \label{fig:s2sites}
\end{figure}

The mesh containing $N_s$ points in the first Brillouin zone is
generated by two vectors $\mathbf{t}_1$ and $\mathbf{t}_2$, whose cell area
satisfies $\lvert \mathbf{t}_1 \times \mathbf{t}_2 \rvert
=\lvert \mathbf{G}_1 \times \mathbf{G}_2 \rvert / N_s$, where $\mathbf{G}_1$ and
$\mathbf{G}_2$ are the primitive reciprocal lattice vectors of the
superlattice. The $N_s$ momentum points of the mesh are then given by
\[
    \mathbf{k}_{ij}=i\,\mathbf{t}_1+j\,\mathbf{t}_2,\qquad
    i=0,\ldots,m-1,\qquad j=0,\ldots,n-1,
\]
with $N_s=m\times n$, so that $N_s$ translated copies of the elementary cell of area
$\lvert\mathbf{t}_1\times\mathbf{t}_2\rvert$ tile the Brillouin zone.
To be compatible with the
periodic boundary conditions, the mesh vectors must be commensurate with the
reciprocal lattice:
\begin{equation}
  m\mathbf{t}_1=\alpha_1\mathbf{G}_1+\alpha_2\mathbf{G}_2,\qquad
  n\mathbf{t}_2=\beta_1\mathbf{G}_1+\beta_2\mathbf{G}_2,
\end{equation}
with integers $m$, $n$, $\alpha_i$, and $\beta_i$.
The commensurability conditions guarantee that shifting a momentum point by $m\mathbf{t}_1$ or $n\mathbf{t}_2$
brings it back to an equivalent point modulo a reciprocal lattice vector, so
that the mesh forms a closed discrete torus and the Bloch states at equivalent
points are identical up to a gauge phase. Figure~\ref{fig:s2sites} shows the meshes with 26 and 28 momentum points.

\subsection{Particle entanglement spectrum}
To obtain an independent diagnostic of the topological order of the
ground-state manifold, we calculate its particle entanglement spectrum
(PES). For an $n_{\rm g}$-fold quasi-degenerate ground-state manifold
$\{|\Psi_i\rangle\}$, we first construct the equal-weight mixed density
matrix
\begin{align}
  \rho_{\rm GS}
  = \frac{1}{n_{\rm g}}
  \sum_{i=1}^{n_{\rm g}}|\Psi_i\rangle\langle\Psi_i|.
\end{align}
We then partition the $N_e$ particles into subsystems containing
$N_A$ and $N_B=N_e-N_A$ particles and trace over subsystem $B$,
\begin{align}
  \rho_A=\operatorname{Tr}_B\rho_{\rm GS}.
\end{align}
If $\lambda_\alpha$ are the eigenvalues of $\rho_A$, the particle
entanglement energies are defined as
$\xi_\alpha=-\ln\lambda_\alpha$, where $\alpha$ labels the entanglement levels.
In a topologically ordered phase, a
well-defined entanglement gap can separate a universal low-lying PES
manifold from higher entanglement levels. The number and momentum-sector
distribution of the levels below this gap can then be compared with the
quasihole counting predicted for a candidate topological order.

\begin{figure}
  \centering
  \includegraphics[width=1\linewidth]{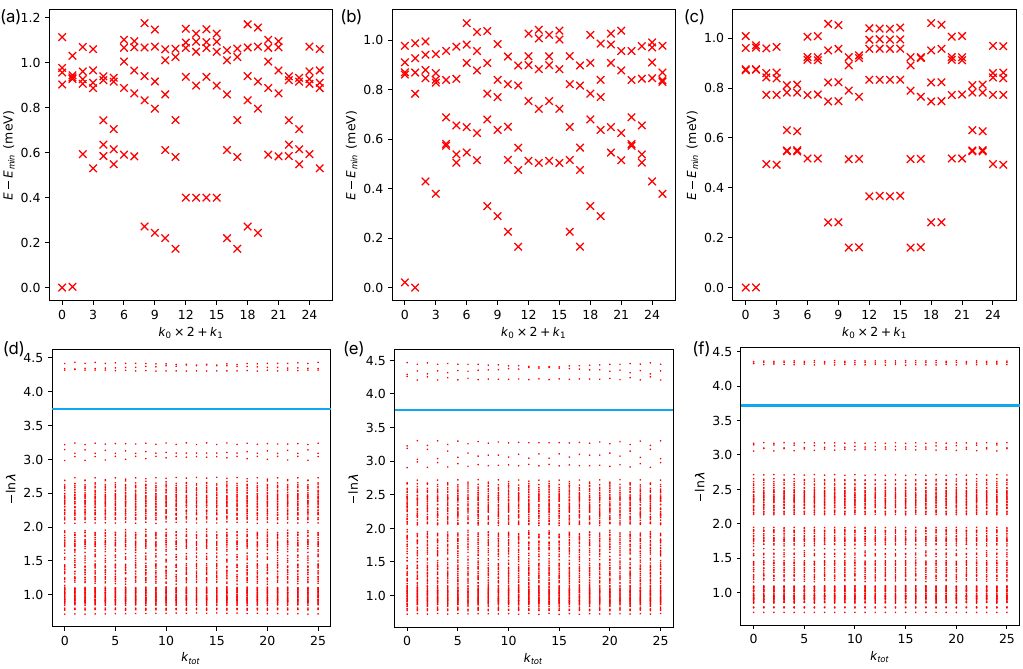}
  \caption{Many-body energy spectra on 26-site momentum meshes for the (a) triangular, (b) honeycomb, and (c) kagome lattices. The twofold quasi-degenerate ground-state manifold is consistent with that expected for the Moore-Read state. The corresponding PES are shown in panels (d)--(f).}
  \label{fig:figs326pess}
\end{figure}

In addition to the 28-site results in the main text, we present the 26-site energy spectra and PES obtained from ED calculations for the triangular, honeycomb, and kagome lattices. Figures~\ref{fig:figs326pess}(a)--(c) present the energy spectra for the three lattices, each showing a twofold quasi-degenerate ground-state manifold with momentum sectors consistent with those expected for a Moore-Read state on the 26-site torus. At half filling, the 26-site mesh accommodates an odd number of electrons, $N_e=13$. For the Moore-Read (Pfaffian) state, the ground-state degeneracy on a torus is six for an even number of electrons and two for an odd number of electrons, with the latter corresponding to the sector containing an unpaired Majorana mode \cite{moore-read,wen-kagome-prl11}. The observed twofold degeneracy is therefore consistent with the odd-particle-number sector of the Moore-Read state.

To further characterize the ground states, we compute the PES of the 26-site ground-state manifolds by partitioning the $N_e=13$ electrons into two subsystems, $A$ and $B$, with $N_A+N_B=N_e$ particles.
For the $N_A=4$ cut, the entanglement gap is rather small, making the identification of the low-lying levels ambiguous.
We therefore compute the $N_A=3$ PES, which exhibits a much larger entanglement gap with 2522 low-lying levels, consistent with the Moore-Read quasihole counting~\cite{fu-mote2-prl24}, as shown in Figs.~\ref{fig:figs326pess}(d)--(f).

\subsection{Chiral-model phase diagrams}

In the ``chiral approximation" of interlayer coupling, one sets the interlayer Fermi velocity $\hbar v_{\perp}\to 0$, reducing the interlayer hopping $h_{\mathrm{inter}}^{0,\mu}$ (Eq.~\eqref{eq:hintra-inter}) to a purely off-diagonal coupling from the $A$ sublattice of one layer to the $B$ sublattice of the other:
\begin{equation}
  h_{\mathrm{inter,chiral}}^{0,\mu}=
  \begin{pmatrix} 0 & t_{\perp} \\ 0 & 0 \end{pmatrix}.
\end{equation}
The superlattice potential term remains unchanged. In such ``chiral model", the Berry curvature distribution simplifies, providing a feasible starting point for studying the superlattice band structures and many-body ground states.

\begin{figure}
	\centering
	\includegraphics[width=\linewidth]{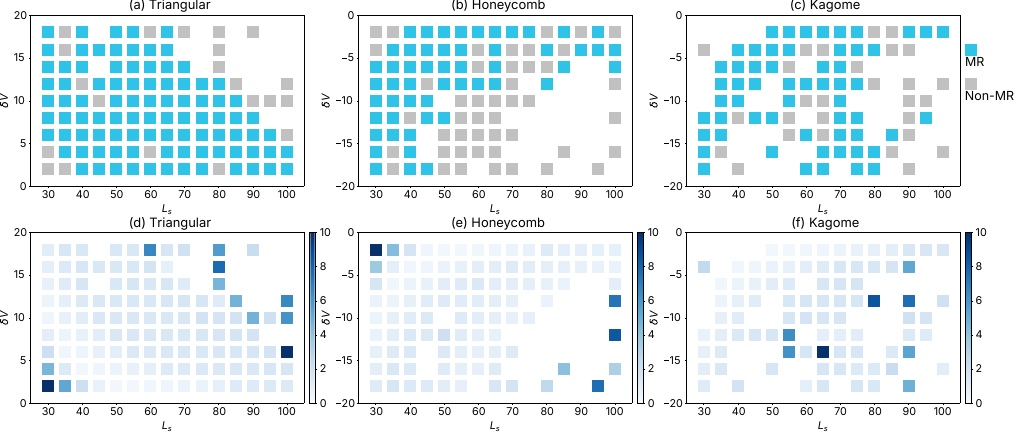}
	\caption{Chiral limit phase diagrams of the HVB at 1/2 filling in the $(L_s,\delta V)$ parameter space for (a) triangular, (b) honeycomb, and (c) kagome patterned superlattices coupled to bilayer graphene. Moore-Read FCI states are colored light blue, the gray blocks denote many-body states that are not Moore-Read FCIs, and the white regions indicate parameter points at which the Chern number of the HVB is not $\pm1$ and the many-body ground state was therefore not calculated. (d)--(f) show the converged values of the loss function for the triangular, honeycomb, and kagome superlattices, respectively.}
	\label{fig:fig4v5}
\end{figure}
In addition to the full BLG continuum model, we also present the results from the model with pure off-diagonal interlayer coupling. 
Fig.~\ref{fig:fig4v5} shows the many-body phase diagram of such chiral BLG model. 
Moore-Read FCI  states emerge over a much larger parameter range than in the full model, and the occurrence of Moore-Read ground states correlates with the low loss function in an evident manner.

\subsection{Scatter plot of Moore-Read FCI distributions}
\begin{figure}
	\centering
	\includegraphics[width=0.5\linewidth]{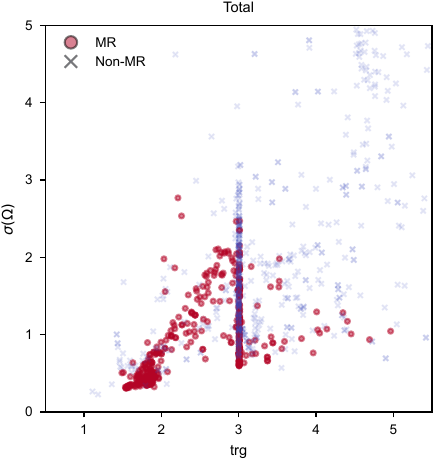}
	\caption{Distribution of MR and other states versus the trace condition (trg) and standard deviation of Berry curvature ($\sigma(\Omega)$). Both the data from chiral and full models are included.}
	\label{fig:total}
\end{figure}
We plot the distribution of Moore-Read FCI and non-Moore-Read states versus the trace of quantum metric and the standard deviation of the Berry curvature in Fig.\ref{fig:total}, including data from both the chiral and the full continuum models of BLG. The horizontal axis shows the trace of quantum metric, $\operatorname{trg} = \int_{\mathbf{k}} \operatorname{tr} g(\mathbf{k})$, which equals to 3 for the ideal first Landau level, and the vertical axis shows the standard deviation of the Berry curvature $\sigma(\Omega)$; these are precisely the two descriptors entering the loss function. 
Red filled circles denote cases for which the half-filled many-body ground state is a Moore-Read FCI. These Moore-Read FCI states are identified by the following criteria: (a) the presence of low-lying quasi-degenerate ground states at momentum sectors consistent with those of Moore-Read FQH state on 28-site torus, with the gap between quasi-degenerate ground states and the excited state larger than the energy spread within the ground-state manifold; (b) with $N_A=3$, the presence of clear PES gap and correct counting of low-lying PES levels below the gap; and (c) with $N_A=4$, the presence of at least a dip in PES at the level position below which the number of low-ying levels are consistent with the corresponding quasi-hole counting. 
Blue crosses in Fig.~\ref{fig:total} denote the non-Moore-Read states, including charge density wave and Fermi liquid etc. 
While the Moore-Read states concentrates within the range of $\operatorname{trg} $ close to 3 and lower $\sigma(\Omega)$, plenty of non-MR states resides in the same parameter range. Some Moore-Read FCI states emerge in other parameter region where the quantum geometry of the target band is not too close that of the first LL, e.g. with $\operatorname{trg}\sim 2$ and $\sigma(\Omega)\lessapprox 0.5$ in the lower left part of the figure.
The two descriptors in the loss function provides an effective indicator for the MR states, while the many-body ground state is still much affected by other parameter such as the band dispersions etc. It implies that Moore-Read FCI seems to have a more complicated mechanism than the corresponding parent FQH state, which deserves further study.

\end{document}